\documentclass[aps,prl,twocolumn,superscriptaddress]{revtex4-1}
\usepackage{amssymb,mathrsfs,mathtools,overpic,physics,tikz}
\usepackage[caption=false]{subfig}

\newcommand{\noise}{\mathcal{E}}
\newcommand{\su}[1]{\text{SU}(#1)}
\DeclareMathOperator{\std}{\text{std}}

\definecolor{tqblue}{RGB}{38,151,208}
\definecolor{tqgreen}{RGB}{66,184,99}

\begin{document}

\title{Simulating and mitigating crosstalk}
\author{Adam Winick}
\affiliation{Quantum Benchmark Inc., 51 Breithaupt Street\\ Suite 100, Kitchener, ON N2H 4C3, Canada}
\affiliation{Institute for Quantum Computing, University of Waterloo, Waterloo, Canada}
\author{Joel J. Wallman}
\affiliation{Quantum Benchmark Inc., 51 Breithaupt Street\\ Suite 100, Kitchener, ON N2H 4C3, Canada}
\affiliation{Institute for Quantum Computing, University of Waterloo, Waterloo, Canada}
\author{Joseph Emerson}
\affiliation{Quantum Benchmark Inc., 51 Breithaupt Street\\ Suite 100, Kitchener, ON N2H 4C3, Canada}
\affiliation{Institute for Quantum Computing, University of Waterloo, Waterloo, Canada}
\affiliation{Canadian Institute for Advanced Research, Toronto, Ontario, Canada}

\begin{abstract}
We describe an efficient and scalable framework for modeling crosstalk effects on quantum information processors. By applying optimal control techniques, we show how to tuneup arbitrary high-fidelity parallel operations on systems with substantial local and nonlocal crosstalk. Simulations show drastically lower error rates for a 2D square array of 100 superconducting transmon qubits. These results suggest that rather than striving to engineer away undesirable interactions during fabrication, we can largely mitigate their effects through careful characterization and control optimization.
\end{abstract}

\maketitle

\emph{Introduction.} --- The foremost obstacle to realizing practical quantum computing is its innate sensitivity to errors and noise. One facet of the problem is the dichotomy between the implementation of high-fidelity simultaneous one-qubit and high-fidelity two-qubit gates. Fast two-qubit gates require spatially or spectrally nearby qubits, which reduces the constituent subsystems' addressability because a resonant pulse intended for one qubit can induce rotations on the others \cite{Gershenfeld1997,Pioro2008}. Proposed architectures have typically dealt with this crosstalk by attempting to maximize the gap between qubits or by executing local operations asynchronously \cite{Schutjens2013,Piltz2013,Boixo2018}. The former solution requires the ability to tune couplings or extra engineering, but the added complexity can adversely impact coherence times and requires additional control wires. In the latter approach, depending on the extent to which the control fields affect neighboring subsystems, the time overhead can be significant.

Crosstalk describes a broad range of effects that violate one of two assumptions: spatial locality and independence of operations \cite{Rudinger2019,Sarovar2019,Abrams2019}. Gates and other operations are supposed to act on disjoint subsets of qubits. However, unintended interactions can couple the qubits, producing nonlocal correlated noise. Even if an operation has a well-defined action on a particular subset of qubits, the effective noise might depend on its context -- what operations affect other qubits.

In this Letter, we introduce a scalable framework for accurately modeling idle and operation crosstalk on experimental devices. Our technique exploits the tensor product structure of local (classical) crosstalk to efficiently express its impact on gates. Through a perturbative expansion, we extend our ideas to nonlocal (quantum) crosstalk and capture its effects to arbitrary order. Provided there is a sufficient degree of control, we can try to minimize the effect of these errors. We illustrate our framework's novel applications through a series of simulations of parallel gates on a square array of 100 superconducting transmon qubits. In our first experiment, we apply gradient-based optimization to the experimentally significant problem of implementing arbitrary elements of $\su{2}^{\otimes n}$ on superconducting transmon qubits. Despite substantial local crosstalk, we show that error rates near the crosstalk-free limit are possible with modern control hardware. We further show how to tuneup simultaneous cross-resonance gates and, again, obtain dramatically higher error rates. Our results suggest that contrary to prevailing opinions \cite{Aude2017,Neill2018,Reagor2018,Murali2020}, crosstalk need not be a prohibitive limitation on noisy intermediate-scale quantum (NISQ) era devices \cite{Preskill2018}. Higher quality quantum information processors may be made possible by using our techniques to better balance the tradeoffs in device fabrication and pulse design.

\emph{Background.} --- Prior work has often approached the problem of implementing several operations on a collection of qubits by breaking it into a temporally disjoint sequence of gates. In contrast, Ref.~\cite{Steffen2000} analyzed the problem of driving two spins with a homogenous field in the setting of NMR. However, it is unclear how to apply the method to multilevel systems such as transmons or trapped ions. Ref.~\cite{Theis2016} studied how to drive two transmons coupled to the same cavity suffering from spectral crowding with simultaneous $X$ or $Y$ gates with rotation angles $\pi$ and $\pi/2$. In either case, these methods do not directly apply to many-qubit systems, nor do they handle nonlocal correlations. It is our objective to develop an efficient and systematic method for optimizing the implementation of nontrivial parallel operations under general crosstalk.

What crosstalk \cite{Sarovar2019} acts on physical qubits during idling or the implementation of gates (as opposed to preparation or measurement crosstalk), and how can we efficiently simulate and consequently, try to mitigate it?  It is natural to classify crosstalk as either local or nonlocal. Local crosstalk can arise when a semiclassical drive field interacts with several qubits, causing unitary errors on supposedly idle qubits, but not entangling independent subsystems. Nonlocal crosstalk creates correlations that are nonfactorizable over system qubits and may originate from, for example, the residual static coupling between two qubits or miscalibration.

Quantifying and reducing crosstalk requires a figure of merit. Depending on the application, it makes sense to evaluate the average fidelity of one-qubit or two-qubit gates rather than the fidelity per clock cycle. Our ideas work in either case, but we focus on the former situation. Local error measures relate directly to fault-tolerance thresholds, are easier to estimate experimentally, and are more common in the literature. We show that the average local fidelity is especially simple to approximate.

\emph{Local crosstalk.} --- Although local crosstalk (typically) produces correlated noise, it can be factorized and simulated efficiently on a digital computer. The induced correlations are classical and do not entangle the individual subsystems. We model local crosstalk via the Hamiltonian
\begin{equation} \label{eqn:LocalHamiltonian}
H(t,\vec x)=\sum_kH_k(t,\vec x) \,.
\end{equation}
Each term $H_k$ acts exclusively on subsystem $k$, and $\vec{x}$ denotes shared classical parameters that result in crosstalk. The vector $\vec{x}$ may, for example, contain the phases and amplitudes that specify drive fields. The average process fidelity $\Phi$ \cite{Nielsen2002b,Carignan2019} between a target operation $U=U_1\otimes\dots\otimes U_n$ and the noisy implementation $\tilde{U}=\tilde{U}_1\otimes\dots\otimes\tilde{U}_n$, where $\tilde{U}_k = \mathcal{T}\exp[-i\int d\tau H_k(\tau, \vec{x})]$, can be expressed as
\begin{equation} \label{eqn:MultiplicativeFidelity}
\Phi(U,\tilde{U}) = \prod_k \Phi(U_k, \tilde{U}_k) \,.
\end{equation}
The equation holds more generally when $\{\tilde{U}_k\}$ are CPTP maps, for example, when a dissipative process also affects the system or the control parameters fluctuate over time.

\emph{Nonlocal crosstalk.} --- Unlike local crosstalk, a digital computer cannot usually exactly simulate a large system affected by nonlocal crosstalk. Thus we develop a perturbative technique for simulating nonlocal crosstalk. Our approximation scheme characterizes a noise channel $\mathcal{E}$ by estimating some of the associated Pauli error rates $\{p_a\}$. The Pauli-twirled noise channel is
\begin{align} \label{eqn:PauliChannel}
	\noise^{\mathcal{P}}(\rho) &= \frac{1}{\abs{\mathcal{P}^n}}\sum_{a\in\mathcal{P}^n}P_a^\dagger\mathcal{E}(P_a\rho P_a^\dagger)P_a \\
	&= \sum_{a\in \mathcal{P}^n}p_aP_a\rho P_a^\dagger \,,
\end{align}
where $\mathcal{P}^n$ is the Pauli group on $n$ qubits. These error rates provide a partial description of the noise affecting a quantum system and can assist in experimental device calibration. On large experimental devices, we can scalably and estimate the parameters in a way that is robust to state preparation and measurement (SPAM) errors \cite{Flammia2019}. We might also combine the quantities to calculate holistic measures of device performance, such as the average two-qubit fidelity or global fidelity.

It is helpful to sketch our approach using a graphical model of the noise (see, e.g., Ref.~\cite{Diestel2005} for basic graph theory definitions). We construct a graph $G$ where each node is a strongly interacting subsystem during an operation of interest, such as a qubit during a single-qubit gate or a two-qubit pair entangled by a cross-resonance interaction. The entire target operation is factorizable over the tensor product space partitioning defined by the nodes. Edges denote nonlocal crosstalk that couples subsystems, and we only allow two-body coupling. We impose the constraint that the graph has limited connectivity (in a spatial  spectral sense) since our approach relies on simulating subsystems. The constraint is satisfied in contemporary architectures where a majority of nodes have a degree of at most four.

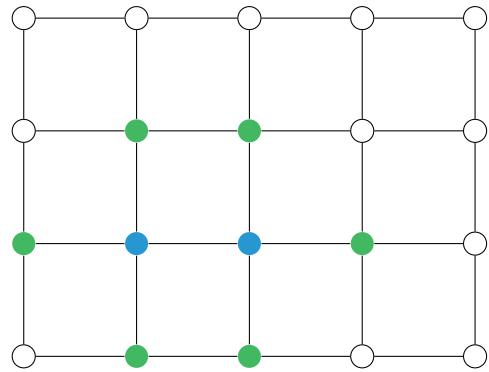
\begin{figure}[t]
\begin{tikzpicture}[font=\sffamily]
%

\node[shape=circle, draw=black] (A1) at (0, 0) {};
\node[shape=circle, fill=tqgreen] (A2) at (1.5, 0) {};
\node[shape=circle, fill=tqgreen] (A3) at (3, 0) {};
\node[shape=circle, draw=black] (A4) at (4.5, 0) {};
\node[shape=circle, draw=black] (A5) at (6, 0) {};
\node[shape=circle, fill=tqgreen] (B1) at (0, 1.5) {};
\node[shape=circle, fill=tqblue] (B2) at (1.5, 1.5) {};
\node[shape=circle, fill=tqblue] (B3) at (3, 1.5) {};
\node[shape=circle, fill=tqgreen] (B4) at (4.5, 1.5) {};
\node[shape=circle, draw=black] (B5) at (6, 1.5) {};
\node[shape=circle, draw=black] (C1) at (0, 3) {};
\node[shape=circle, fill=tqgreen] (C2) at (1.5, 3) {};
\node[shape=circle, fill=tqgreen] (C3) at (3, 3) {};
\node[shape=circle, draw=black] (C4) at (4.5, 3) {};
\node[shape=circle, draw=black] (C5) at (6, 3) {};
\node[shape=circle, draw=black] (D1) at (0, 4.5) {};
\node[shape=circle, draw=black] (D2) at (1.5, 4.5) {};
\node[shape=circle, draw=black] (D3) at (3, 4.5) {};
\node[shape=circle, draw=black] (D4) at (4.5, 4.5) {};
\node[shape=circle, draw=black] (D5) at (6, 4.5) {};

\path[-, draw=black] (A1) edge (A2);
\path[-, draw=black] (A2) edge (A3);
\path[-, draw=black] (A3) edge (A4);
\path[-, draw=black] (A4) edge (A5);
\path[-, draw=black] (B1) edge (B2);
\path[-, draw=black] (B2) edge (B3);
\path[-, draw=black] (B3) edge (B4);
\path[-, draw=black] (B4) edge (B5);
\path[-, draw=black] (C1) edge (C2);
\path[-, draw=black] (C2) edge (C3);
\path[-, draw=black] (C3) edge (C4);
\path[-, draw=black] (C4) edge (C5);
\path[-, draw=black] (D1) edge (D2);
\path[-, draw=black] (D2) edge (D3);
\path[-, draw=black] (D3) edge (D4);
\path[-, draw=black] (D4) edge (D5);

\path[-, draw=black] (A1) edge (B1);
\path[-, draw=black] (A2) edge (B2);
\path[-, draw=black] (A3) edge (B3);
\path[-, draw=black] (A4) edge (B4);
\path[-, draw=black] (A5) edge (B5);
\path[-, draw=black] (B1) edge (C1);
\path[-, draw=black] (B2) edge (C2);
\path[-, draw=black] (B3) edge (C3);
\path[-, draw=black] (B4) edge (C4);
\path[-, draw=black] (B5) edge (C5);
\path[-, draw=black] (C1) edge (D1);
\path[-, draw=black] (C2) edge (D2);
\path[-, draw=black] (C3) edge (D3);
\path[-, draw=black] (C4) edge (D4);
\path[-, draw=black] (C5) edge (D5);
\end{tikzpicture}
\caption{Graphical model depicting a system of 20 qubits on a grid with nearest-neighbor nonlocal crosstalk. We highlight 1 of 31 (there are 31 edges; one for each nonlocal interaction) noise simulations needed to estimate all marginal weight-2 Pauli error probabilities with the approximation $(d,o)=(1,2)$.  The blue and green nodes denote the target subsystem and the environment.}
\label{fig:NonlocalExpansion}
\end{figure}

A pair of positive integers $(d, o)$ specifies the expansion order of the noise approximation; $d$ designates the `environment' distance and $o$ the maximum component order. We consider the set $\mathcal{G}_o$ of all components of all induced subgraphs of $G$ such that the order of every component is less than or equal to $o$, and any component with an order less than $o$ has the same edges as in $G$. I.e., we do not look at induced components with order less than $o$. The idea of the simulation scheme is to calculate the Pauli errors that occur on each component. 

We approximate the behavior of a component $C \in \mathcal{G}_o$ by evolving it along with all vertices of distance at most $d$, generating a map $\noise_{C,d}$. Next, we compute the diagonal of the Pauli-Liouville representation of the channel $f_{C,d}$. A Walsh-Hadamard transformation $W$ relates $f_{c,d}$ to the Pauli probability vector $\tilde p_{C,d}$, with $f_{C,d} = W\tilde p_{C,d}$ \cite{Flammia2019}. The vector $\tilde p_{C,d}$ is the error probability distribution for a Pauli twirled copy of $\noise_{C,d}$. Marginalizing the error distribution over the environment produces an estimate of the local error distribution $\tilde p_C$ on the target component. After calculating the marginal distributions for all of the components in $\mathcal{G}_o$, we can use the theory of probabilistic graphical models \cite{Koller2009} to construct an estimate of the entire Pauli error distribution up to some specified error weight. By truncating the distribution at some error weight, the size of the distribution scales polynomially in the number of qubits.

In practice, including the nearest environmental nodes is sufficient to compute the local error distribution with high relative precision. We can intuitively understand the limited-depth requirement from the fact that intermediate systems must mediate the influence of one subsystem on a nonadjacent subsystem. One may formally bound these effects with Lieb-Robinson bounds \cite{Lieb1972}.

\emph{Single-qubit gate engineering.} --- We review a typical implementation of single-qubit operations on transmons. A local oscillator acts as a single tone microwave source outputting a constant signal $\cos(\omega' t)$ that is shaped by an arbitrary waveform generator via an IQ mixer. A good description of a transmon qubit is an anharmonic oscillator driven by microwave pulses. In the lab frame, the relevant Hamiltonian is
\begin{equation} \label{eqn:TransmonHamiltonian}
	H = \omega\hat n+\frac{\alpha}{2}(\hat n - 1)\hat n + \Omega(t)\cos(\omega' t+\gamma)(\hat a + \hat a^\dagger)
\end{equation}
where $\hat a$ is the annihilation operator of the oscillator, $\hat n=\hat a^\dagger \hat a$, $\alpha$ is the anharmonicity, $\gamma$ is the drive phase, $\omega$ is the oscillator's resonant frequency, $\Omega(t)$ specifies the drive envelope, and we set $\hbar = 1$.

The lowest two energy levels form the qubit subspace. Including the third energy level models the leading-order effect of leakage provided the anharmonicity is sufficiently large. After making a rotating wave approximation (RWA) and moving into the rotating frame of the qubit, the Hamiltonian projected into the qubit subspace is
\begin{equation} \label{eqn:RWAQubitHamiltonian}
	H = \frac{1}{2}\Omega(t)e^{-i[\gamma+(\omega'-\omega)t]}\ketbra{0}{1} + \text{h.c.}
\end{equation}
To see how the control induces single-qubit gates, consider a resonant pulse ($\omega = \omega'$), and $\lambda = 0$, which corresponds to an ideal sufficiently long pulse. The control generates $X$ and $Y$ gates by modulating the coupling between the zero and one states, while the drive phase fixes the rotation axis in the $XY$-plane, and the pulse area sets the rotation angle. Rotations about the remaining $Z$-axis correspond to a change in the relative phase between the states. Rather than manipulating the transmon's state, it is equivalent to rotate the control with respect to the state, realizing a virtual-$Z$ gate \cite{Knill2000c,Knill2008,McKay2017b}. We accomplish this physically by adding a phase offset to all subsequent gates. A pulse with an area $\int dt \Omega(t)=\pi/2$ and a relative phase offset $\gamma$ generates the unitary $V(\gamma)=Z_{-\gamma}X_{\pi/2}Z_\gamma$ with the notation $A_\theta=\exp(-i\theta A/2)$. Combining two of these phase-offset $\pi/2$ pulses and a final virtual-$Z$ realizes any element of $\su{2}$ \cite{McKay2017b}.

\begin{figure}[t]
\begin{flushright}
\begin{overpic}[width=.95\linewidth]{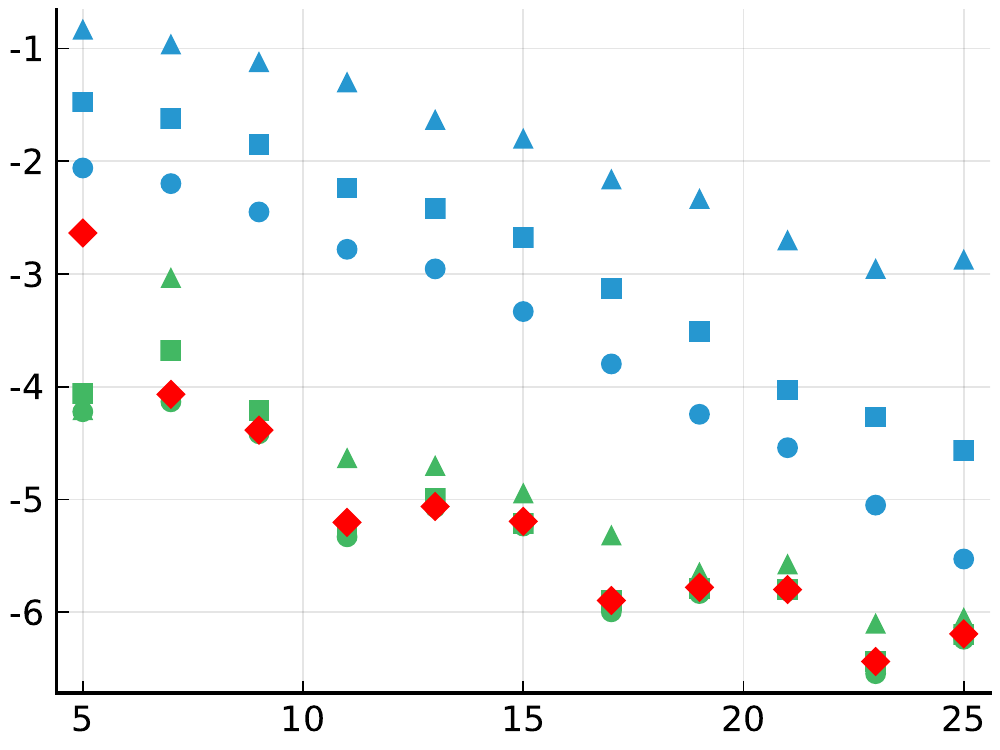}
	\put(48,-4){$t_{\pi/2}$ (ns)}
	\put(-4,34){\rotatebox{90}{$\log_{10}r_\text{avg}$}}
\end{overpic}
\end{flushright}
\caption{Plots illustrating a significant improvement in the average single-qubit process infidelity as a function of the time for a $\pi/2$ gate (the total simulation time is $2t_{\pi/2}$). There are 100 qubits in a square 2D array, and each qubit implements a random element of $\text{SU}(2)$ via two $\pi/2$ pulses with intermediate phase offsets. The red diamonds denote the infidelity of the qubits with half-derivative DRAG corrections and no crosstalk. The blue points are infidelities obtained under the same control with crosstalk. Circles, squares, and triangles denote several relative crosstalk strengths $\beta_{ij}$, that are sampled from a normal distribution with mean zero and standard deviations $\sigma=0.05$, $\sigma=0.1$, and $\sigma=0.25$ respectively. The green markers have identical crosstalk as their blue counterpart, but with optimized control parameters.}
\label{fig:LargeSingleQubit}
\end{figure}

Consider the problem of implementing an arbitrary element of $\su{2}^{\otimes n}$ concurrently on an ensemble of qubits where their respective drive fields weakly interact with other qubits. The semiclassical Hamiltonian governing transmon $k$ with local drive crosstalk is
\begin{equation} \label{eqn:LocalCrosstalkTransmon}
\begin{split}
	H_k &= \omega_k\hat n+\frac{\alpha_k}{2}(\hat n-1)\hat n\\
	&+ \sum_j\beta_{jk}\Omega_j(t)\cos(\omega_j' t+\phi_j+\theta_{jk})(\hat a+\hat a^\dagger) \,.
	\end{split}
\end{equation}
The parameters $\beta$ and $\theta$ characterize the crosstalk affecting the system. We focus on the case where each transmon has a local drive. The crosstalk parameters are $n\times n$ matrices, and we can set $\beta_{kk}$ = 1 and $\theta_{kk}$ = 0 without loss of generality by modifying $\Omega_k$ and $\phi_k$. These constraints lead us to interpret $\beta$ as the relative drive strength, and $\theta$ as the phase lag. Experimental data supports the model \cite{Xia2015,Magesan2018,Xue2019}, and one can efficiently estimate the parameters with standard Rabi and Ramsey experiments.

We simulate a system of $n=100$ transmons that evolve under \eqref{eqn:LocalCrosstalkTransmon} and include the first three energy levels. The qubits are on a square grid with $\beta_{jk}$ nonzero only for neighboring qubits. Qubits have a frequency of either $\omega/2\pi=3$ GHz or $\omega/2\pi=3.1$ GHz, and no two adjacent qubits have the same frequency. All qubits have anharmonicity $\alpha/2\pi = -330$ MHz. In each iteration of the experiment, the target gate is chosen randomly from $\su{2}^{\otimes n}$. The crosstalk phase lag parameter $\theta_{jk}$ are sampled randomly from the interval $[0,2\pi)$, and we draw $\beta_{jk}$ from a normal distribution centered at zero. There are two discrete periods of successive evolution, each taking time $t_{\pi/2}$. It is necessary to pick pulse shapes. On the one hand, we want pulses that yield error rates near the decoherence limit for short gate times. On the other hand, there are experimental realities, such as power-bandwidth constraints and the degree of calibration needed to implement complicated pulses accurately. Balancing these constraints, we pick Gaussian pulses with $\std\Omega^{(x)}=t_{\pi/2}/4$, and half-derivative DRAG corrections $\Omega^{(y)}=-\dot\Omega^{(x)}/2\alpha$ \cite{Motzoi2009,Gambetta2011,Chen2016}.

Fig.~\ref{fig:LargeSingleQubit} shows the average single-qubit process infidelity $r_\text{avg}=1-\expval{\Phi_k}$ as a function of $t_{\pi/2}$. Green diamonds denote the raw infidelity for a crosstalk-free system ($\beta_{jk} = \delta_{jk}$, where $\delta_{jk}$ is the Kronecker delta). The blue markers are infidelities obtained using the crosstalk-free control scheme but with various strengths of drive crosstalk. The red markers are infidelities obtained with optimized control and the same drive crosstalk as the blue markers. We optimize control pulses with the method of Ref.~\cite{Machnes2018}. Applying the protocol requires the selection of appropriate optimization parameters. Sticking to our simple control ansatz, we tune the overall magnitude of the resonant $\Omega^{(x)}$ quadrature, off-resonant $\Omega^{(y)}$ quadrature and the carrier signal phase $\phi$, for a total of $7n$ parameters. We observe approximately two orders of magnitude improvement in the infidelity with our crosstalk minimization technique.

In real experimental devices, decoherence significantly reduces the average error rates. Moreover, decoherence errors grow with time, whereas control errors typically decrease. These contrasting effects imply that there is an optimal gate time that minimizes their combined errors. We repeat the simulation implementing $\su{2}^{\otimes n}$ with decoherence added to the model. Table~\ref{table:su2decoh} presents data showing the potential benefit of our methods.

\begin{table}[t]
\begin{center}
\begin{tabular}{l|c|c|c}
	Crosstalk & Original & Opt., $t_{\pi/2}=2$ ns & Opt., $t_{\pi/2}=5$ ns\\
	$\std\beta_{jk}$ & $r_\text{avg}$ & $r_\text{avg}$ & $r_\text{avg}$ \\
	\hline
	0.05 & 6.02e-4 & 1.00e-4 & 1.86e-4\\
	0.1 & 7.13e-4 & 1.03e-4 & 1.91e-4\\
	0.25 & 2.13e-3 & 1.15e-4 & 1.84e-4\\
	0.5 & 1.77e-2 & 1.07e-4 & 1.81e-4
\end{tabular}
\caption{Data highlighting a dramatic reduction in the average single-qubit process infidelity for a simulation with realistic decoherence on a square array of 100 qubits and various levels of crosstalk ($\std\beta_{jk}$). We model the same system considered in Fig.~\ref{fig:LargeSingleQubit}, but with $T_1\sim\mathcal{N}(40\text{ $\mu$s}, 5\text{ $\mu$s})$ for each qubit, and $T_2=3T_1/2$.
Naturally, there is an optimal gate time that minimizes the combined incoherent (increasing) and coherent (approximately decreasing) effects. We optimize the controls for $t_{\pi/2} = 1, 2, \dots, 50$ ns. The `Original' column corresponds to the optimal $t_{\pi/2}$ without control tuneup. For all values of $\std\beta_{j,k}$, $r_\text{avg}$ is minimized at $t_{\pi/2}=2$ ns. On contemporary experimental devices, $t_{\pi/2}=2$ ns exceeds accesible bandwidths, so we also report $r_\text{avg}$ for $t_{\pi/2}=5$ ns.} \label{table:su2decoh}
\end{center}
\end{table}

\emph{Two-qubit gate engineering.} --- We continue our simulations using the ideal system of fixed-frequency transmons and the parameter values specified above. Our aim is to implement parallel cross-resonance gates \cite{Rigetti2005,Chow2011,Sheldon2016,Magesan2018,Allen2019}, which are equivalent to CNOTs up to single-qubit operations. Constant capacitive coupling provides a mechanism for implementing entangling operations. Assuming equal coupling between all neighboring qubits in the system, the corresponding interaction Hamiltonian is
\begin{equation} \label{eqn:InteractionHamiltonian}
	H_\text{int}=J\sum_{\langle j,k\rangle =1} a_j a_k^\dagger + a_j^\dagger a_k \,,
\end{equation}
where $\langle j,k\rangle =1$ denotes a sum over all adjacent qubit pairs. The entire system evolves under $H_\text{int}+\sum_k H_k$.

The basic idea of the cross resonance effect is that if we define the qubits in a dressed basis, local microwave drive fields drive both single and two-qubit gates. For two ideal coupled qubits, in the dressed basis, a drive applied to qubit 1 at the frequency of qubit 2 yields the effective Hamiltonian
\begin{equation}
	H_d=\Omega(t)\left(X_1-\frac{J}{\Delta}Z_1X_2\right) \,,
\end{equation}
where $\Delta=\omega_1-\omega_2$ is the difference of qubit frequencies and we made an RWA. Although we can decouple the direct qubit coupling, higher-levels of the transmon lead to additional terms in the effective Hamiltonian \cite{Magesan2018}. We can use the $Z_1X_2$ term to generate a maximally entangling gate.

\begin{figure}[t]
\begin{flushright}
\begin{overpic}[width=.95\linewidth]{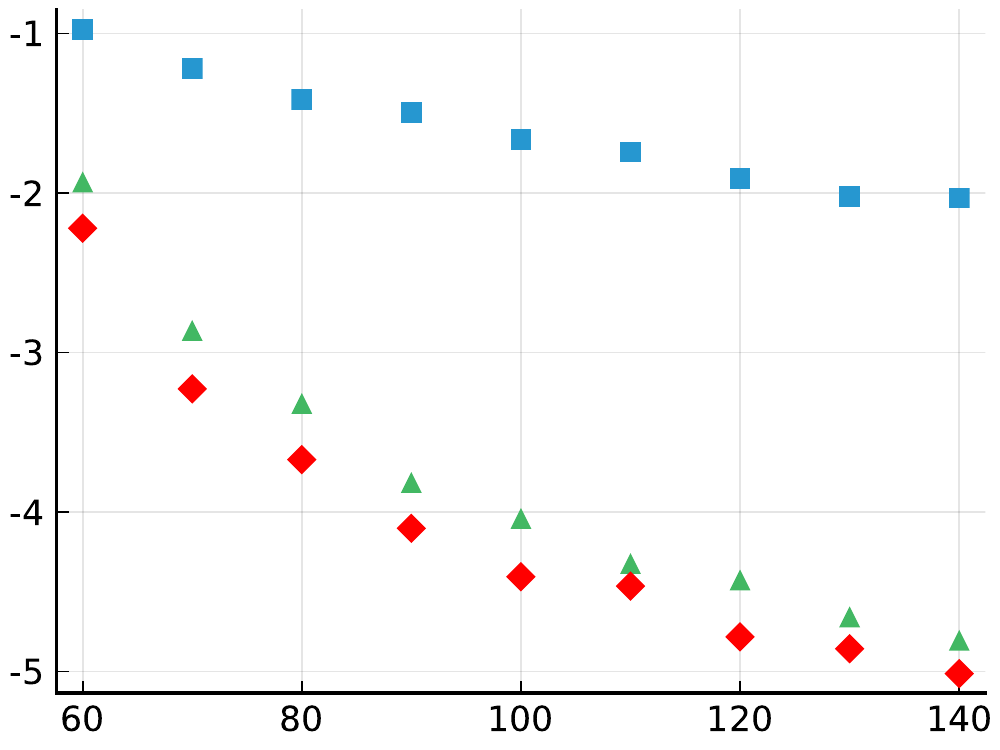}
	\put(48,-4){$t_\text{CR}$ (ns)}
	\put(-4,26){\rotatebox{90}{$\log_{10}\expval{1-\Phi_\text{CR}}$}}
\end{overpic}
\end{flushright}
\caption{Plots showing a massive improvement in the average two-qubit process infidelity for a square array of transmons implementing 50 simultaneous maximally entangling gates via CR interactions. Each two-qubit pair approximates a CNOT-gate up to local operations. The red diamonds correspond to CR gate infidelities obtained without drive crosstalk or subsystem coupling. The blue squares are infidelities obtained with the same controls as the green diamonds but with drive crosstalk as in the single-qubit example ($\sigma=0.1$) and constant nonlocal coupling between all adjacent qubits. The green triangles have the same crosstalk as the above model, but with optimized control parameters.}
\label{fig:LargeTwoQubit}
\end{figure}

Again, we simulate a system of $n=100$ transmons on a grid and include the first three energy levels of each. We group adjacent qubits in pairs and try to implement 50 simultaneous maximally entangling gates using the CR effect. Our qubits have 8 distinct frequencies ($3.0, 3.1, \dots, 3.7$ GHz) to ensure each CR pair is addressable. We set the frequencies so that no two neighbors of one qubit have the same frequency. The target CNOT equivalent is determined using Cartan's KAK decomposition \cite{Tucci2005} and is invariant to local operations. The qubit coupling strength is $J/2\pi=3.8$ MHz. We realize qubit control with the same drives as above but with variable drive detuning and phase offset. We independently parameterize the resonant $\Omega^{(x)}$ and off-resonant $\Omega^{(y)}$ control envelopes with the first three Hanning window functions
\begin{equation} \label{eqn:HanningWindow}
	\Omega_H(t) = \sum_{k=1}^3 c_k\left[1-\cos(\frac{2\pi k t}{t_\text{CR}})\right] \,.
\end{equation}
There are a total of $8n$ parameters that determine the $n$ drive fields.

Fig.~\ref{fig:LargeTwoQubit} shows the average two-qubit process infidelity of each entangling gate as a function of the gate duration $t_\text{CR}$. We compute all points with optimized pulse parameters \cite{Machnes2018} but under different system models. Red diamonds denote the infidelity obtained using a drive-crosstalk-free model and no undesirable $J$ coupling. The blue squares are infidelities calculated using the crosstalk-free optimal control but with added drive crosstalk ($\sigma=0.1$) and nonlocal coupling. The green triangles are infidelities obtained with controls tuned up under the crosstalk model. We approximate the nonlocal crosstalk effects with $d=1$. The deviation caused by including additional neighbors is unresolvable on the plot.

\emph{Conclusion.} --- We have described techniques to efficiently model and minimize crosstalk that occurs during qubit idling and gates. Compared to other quantum control methods such as dynamical decoupling \cite{Viola1999}, which attempts to echo out undesirable interactions, we change parameters so the effects do not appear in the first place. Our results show how to mitigate such effects on transmons using a fast control tuneup procedure on a digital computer. We hope that these methods aid in understanding the role of crosstalk on NISQ devices and validate improved pulse shapes. We plan to extend our simulation capabilities to other platforms such as trapped ions and apply our methods to improve the performance of experimental platforms.

\emph{Acknowledgments.} --- We thank Ian Hincks and Dar Dahlen for helpful discussions and feedback on the manuscript. This research was supported by the U.S. Army Research Office through grant W911NF-14-1-0103 and Quantum Benchmark Inc.

\bibliography{references}
\bibliographystyle{apsrev4-1}

\end{document}